\documentclass[preprint,12pt]{elsarticle}

\usepackage{bm,amsmath,amssymb,latexsym,mathrsfs,graphicx,enumerate}
\usepackage[mathcal]{euscript}
\usepackage{hyperref}
\usepackage{epsfig}

\graphicspath{{figs/}}
\RequirePackage{graphicx}
\journal{Optics Communications}
\begin{document}
\begin{frontmatter}

\title{\textbf{Nonlinear response of a thin  metamaterial film containing  Josephson junctions }}

\author{Andrei I. Maimistov }
\address{Department of Solid State Physics and Nanosystems, Moscow Engineering Physics Institute, Kashirskoe sh. 31, Moscow 115409, Russia}
\author{Ildar\,R.\,Gabitov }
\address{Department of Mathematics, University of Arizona, 617 North Santa Rita Avenue, Tucson, AZ 85721, USA}

\begin{abstract}
An interaction of electromagnetic field with metamaterial thin film containing split-ring resonators with Josephson junctions is considered.  It is shown that  dynamical self-inductance in a split rings results in reduction of  magnetic flux through a ring and this reduction is  proportional to a time derivative of split ring magnetization. Evolution  of thin film magnetization taking into account dynamical self-inductance is studied. New mechanism for  excitation of   waves  in one dimensional  array of split-ring resonators with Josephson junctions is proposed.  Nonlinear magnetic susceptibility of such thin films is obtained in the  weak amplitude approximation.

\end{abstract}


\end{frontmatter}

\section{Introduction}
\noindent During the last decade metamaterials have become the focus of intensive research~\cite{Pendry:04,Veselago:06,Agranovich:06}. The nonlinear
electrodynamics of metamaterials is of special interest due to the presence of new nonlinear phenomena which are specific to
metamaterials~\cite{Maim:Gabi:07,Maim:Gabi:Lit:08,Kaz:Maim:Ozh:09} and due to the fact that, in several cases, nonlinearity is a
characteristic feature of nanoscale systems~\cite{Rautian:97,Manfredi:97}. The nonlinear response of metamaterials can also be the result of deliberate design.   For example, nonlinear dielectrics or diodes can be inserted into split rings~\cite{Lapine:03,Shadrivov:06}. Josephson junctions (JJ) are known to be strongly nonlinear~\cite{Hansen:84} with low losses and therefore they are a  natural way to introduce nonlinearity in metamaterials. Electromagnetic field interaction with metamaterials containing split-rings with Josephson junctions inserted into the gap was recently considered in several papers. For example, localized oscillations in chains and two dimensional arrays of such split-rings were considered in~\cite{Lazarides:06,Lazarides:07,Lazarides:09}. The existence of metastable states in a medium of this type and transitions between these states was discussed in~\cite{GabMaimSPIE08}.

Currently metamaterials primarily exist in the form of thin films. Thus, it is a clear choice  to study the interaction of an electromagnetic wave with thin films containing strongly nonlinear structural units. Split-rings containing Josephson junctions are natural building blocks for such thin films.  This subject is considered in the first section of this paper. In particular, we investigated the effect of dynamical self-inductance in split rings which results in reduction of magnetic flux through a ring. This reduction is proportional to the time derivative of split ring magnetization. This effect manifests itself as non Fresnellian reflection from a film and represents an additional mechanism increasing rate of oscillation relaxation in ring-resonators. We show that this additional damping leads to dramatic change in the evolution dynamics of film magnetization.

The interaction of a chain of split-rings containing  JJ with an electromagnetic field is considered in the second section. We expressed the individual current of a particular ring in terms of magnetic fluxes through all other rings in a chain. In continuous limit, taking into account near neighbor interactions to leading order, we obtained sine-Gordon  type of equation describing "continuous" chain dynamics. The effect of dynamical self inductance is presented in this model as an additional damping term. External force in this equation contains the second spatial derivatives of the incident field and therefore suggests a new mechanism for the excitation of  longitudinal oscillations by a normally incident field (without tangential component along the chain).

In the third section, we obtained the nonlinear magnetic susceptibility of a thin film containing  split-rings with  JJ in the limit of weak amplitudes.  This susceptibility describes third harmonic generation and self modulation due to high frequency Kerr effect.

\section{Electromagnetic wave  interaction with thin films containing  Josephson junctions
}\label{SSR:JJ}

\noindent Electromagnetic field interaction with thin films ($\Delta
x \ll \lambda$) is a well studied subject in the
literature~\cite{Rupasov:82,Rupasov:87,BBE:86,Bash:88,BMTZ91,Vanagas:98,Elyutin07a,Caputo:07}.
Most of these works consider thin films being polarized under an external
field. In this paper we study magnetoactive thin films. This requires the
derivation of new modeling equations which we  present in the
following subsection.

\subsection{Transmission and reflection of electromagnetic wave on magnetoactive thin films: basic equations}

We consider a plane electromagnetic wave normally incident from
$-\infty$ along the $x$ axis on a magnetoactive thin film.   The
corresponding Maxwell equations have following form:
\begin{equation}
\frac{\partial H_y}{\partial x}= \frac{1}{c}\frac{\partial
D_z}{\partial t},~~ \frac{\partial E_z}{\partial x}=
\frac{1}{c}\frac{\partial B_y}{\partial t}. \label{SRR:JJ:1}
\end{equation}
We take into account only  magnetic response.  Total magnetic inductance reads as
\[
B_y (x,t) = B^{host}(x,t)+4\pi M(t) \delta (x).
\]
Here  magnetization of the thin film $\approx l M^{(s)}$ can be
represented as a product of surface magnetization $M^{(s)}$  and the
width of the  film $l$. $B^{host}(x,t)$ is the magnetic inductance of the
host material. Integrating the equations (\ref{SRR:JJ:1})
over $x$ from $-\delta x$ to $+\delta x$ and taking the limit $\delta x
\rightarrow 0$ leads to the boundary conditions at the point $x=0$:
\begin{eqnarray}
H_y(0-) &=& H_y(0+), \label{SRR:JJ:2} \\
E_z (0-)-E_z(0+) &=& - \frac{4\pi l}{c}\frac{\partial
M^{(s)}}{\partial t}. \label{SRR:JJ:3}
\end{eqnarray}

In a homogeneous medium for $x<0$ and at $x>0$ the system of equations~(\ref{SRR:JJ:1}) can be
reduced to the  wave equation
\begin{eqnarray}
\frac{\partial^2 H}{\partial x^2}=
\frac{\varepsilon}{c}\frac{\partial^2 H}{\partial t^2}\label{Wave:equation:H}
\end{eqnarray}
where $\varepsilon=\varepsilon_1$ for $x<0$ and
$\varepsilon=\varepsilon_2$ for $x>0$. Introducing variables  $q_j =
k_0 \sqrt{\varepsilon_j}, \, j=1, 2$  and $k_0=\omega/c$ we can
represent the solution of~(\ref{Wave:equation:H}) in terms of Fourier
components as follows:
\begin{equation}
\tilde{H}(x,\omega)=\left\{
\begin{array}{lr}
A e^{iq_1 x} + B e^{-iq_1 x}, & x<0, \\
C e^{iq_2 x}, & x>0,
\end{array} \right.
\label{SRR:JJ:4}
\end{equation}
For the electric field we have
\begin{equation}
\tilde{E}_z(x,\omega)=\left\{
\begin{array}{lr}
-q_1 (\varepsilon_1 k_0)^{-1}\left( A e^{iq_1 x} - B e^{-iq_1 x}\right), & x<0, \\
-q_2 (\varepsilon_2 k_0)^{-1}C e^{iq_2 x}, & x>0,
\end{array} \right.
\label{SRR:JJ:5}
\end{equation}

Taking into account boundary conditions  (\ref{SRR:JJ:2}) and
(\ref{SRR:JJ:3}) we obtain relations
\begin{eqnarray*}
A+B &=&C, \\
\frac{iq_1}{\varepsilon_1}(A-B) &=& \frac{iq_2}{\varepsilon_2} C +
4\pi k_{0}^{2}l \tilde{M}^{(s)},
\end{eqnarray*}
which  define the amplitudes $B, C$ via the amplitude of the incident wave $A= \tilde{H}^{in}$:
\begin{equation}
\tilde{H}^{(tr)}=C=\frac{2 \varepsilon_2 q_1}{\varepsilon_2 q_1 +
\varepsilon_1 q_2 }\tilde{H}^{(in)} + \frac{4i\pi k_{0}^{2}l
\varepsilon_1 \varepsilon_2}{\varepsilon_2 q_1 +\varepsilon_1 q_2}
\tilde{M}^{(s)}, \label{SRR:JJ:6a}
\end{equation}
\begin{equation}
\tilde{H}^{(ref)}=B=\frac{\varepsilon_2 q_1 - \varepsilon_1
q_2}{\varepsilon_2 q_1 + \varepsilon_1 q_2 }\tilde{H}^{(in)} +
\frac{4i\pi k_{0}^{2}l \varepsilon_1 \varepsilon_2}{\varepsilon_2 q_1 +
\varepsilon_1 q_2} \tilde{M}^{(s)}. \label{SRR:JJ:6b}
\end{equation}

In the simplest case when $\varepsilon_1 =\varepsilon_2=\varepsilon
$ and dispersion of the host material is negligible ($\varepsilon = const$), expression
(\ref{SRR:JJ:6a}) can be rewritten as
\begin{equation*}
\tilde{H}^{(tr)}= \tilde{H}^{(in)} + \frac{2i\pi l\omega
\sqrt{\varepsilon }}{c} \tilde{M}^{(s)}.
\end{equation*}
In temporal and spatial variables it reads as
\begin{equation}
H^{tr}(t)= H^{(in)}(t) - \frac{2\pi l\sqrt{\varepsilon} }{c}
\frac{\partial M^{(s)}}{\partial t}. \label{SRR:JJ:6amod}
\end{equation}
The expression  for magnetic field inside the film $H(t)=H^{(tr)}$
follows from the continuity condition for  the tangential components
of the magnetic field~(\ref{SRR:JJ:2}). For further analysis we need to
determine properties of the magnetization $M^{(s)}$ for the film
containing Josephson junctions.

\subsection{Magnetic response of split-rings with Josephson junctions}

\begin{figure}[h!]
  \centering \includegraphics[width=1in]{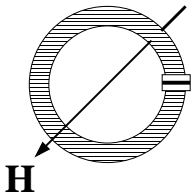}
  \caption{Schematic of a split ring with Josepson junction in the gap. Magnetic field $\textbf{H}$ is orthogonal to the plane of a split ring.}\label{split:ring:JJ}
\end{figure}

\begin{figure}[h!]
  \centering \includegraphics[width=2in]{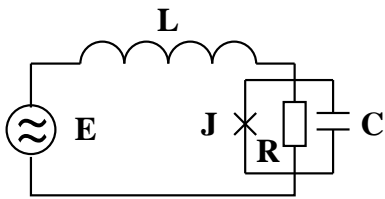}
    \caption{Equivalent electric circuit of a ring with Josephson junction. Here $E$ stands for electromotive force, $L$ is inductance  of a split ring,  $C$ and $R$ are capacitance and resistance of a Josephson junction. }\label{split:ring:JJ:schematic}
\end{figure}

We consider a thin film composed of split rings with Josephson
junctions in their gaps (see Fig.~\ref{split:ring:JJ}). Orientation
of the magnetic field is orthogonal to the split ring's plane.
Fig.~\ref{split:ring:JJ:schematic} shows an equivalent electric circuit
of a split ring with Josephson junction~\cite{Scott:70}. Here
$\mathcal{E}$ stands for electromotive force, $L$ is inductance  of
a split ring,  $C$ and $R$ are capacitance and resistance of a
Josephson junction. The electromotive force connected with the magnetic
flux $\Phi$ is in accordance with Faraday's law:
\begin{equation}
\mathcal{E}=-\frac{1}{c}\frac{d \Phi}{d t}.
\label{electromotive:force}
\end{equation}
Self-induced electromotive force inductance $U_L$ can be expressed
via current $J$ through inductance $L$:
\begin{equation}
U_L = \frac{L}{c^2} \frac{d J}{d t}.\label{self:ind:emf}
\end{equation}
Josephson voltage $U_C$ and current $J_j$ are defined by following
expressions~\cite{Barone:82,Lifscits:Pita:78}:
\begin{equation}
U_C = \frac{\hbar}{2 e}\frac{d \phi}{d t},~~~J_{j}=J_{0}\sin \phi \label{voltage:current:JJ}
\end{equation}
In accordance to Kirchhoff laws the sum of the electrical potential
differences in a circuit is equal to an electromotive force
\begin{equation}
\mathcal{E} = U_L +U_C,  \label{Kirchhoff1}
\end{equation}
and the  sum of currents flowing towards that point is equal to the
sum of currents flowing away from that point, i.e.
\begin{equation}
J_j + J_R + J_C = J. \label{Kirchhoff2}
\end{equation}

The equation~(\ref{Kirchhoff1}) gives
\begin{equation}
\frac{1}{c}\Phi + \frac{L}{c^2} J + \frac{\hbar}{2 e}\phi =0
\label{Integration:Kirchhoff1}
\end{equation}
As follows from~(\ref{Integration:Kirchhoff1}) the current $J$ can
be expressed in the following form
\begin{equation}
J=-\frac{c}{L}\left(\Phi + \frac{\hbar c}{2 e}\phi \right)
\label{Current:J}
\end{equation}
 Let us consider the
second Kirchhoff equation~(\ref{Kirchhoff2}). Variables in the
equation~(\ref{Kirchhoff2}) are defined as follows
\begin{eqnarray*}
J_C &=& C \frac{d U_C}{d t}=C\frac{\hbar}{2 e }\frac{d^{2} \phi}{d t^{2}},\\
J_R &=&\frac{U_C}{R}= \frac{\hbar}{2 e R}\frac{d \phi}{d t}
\end{eqnarray*}
The second Kirchhoff equation~(\ref{Kirchhoff2}) transforms to
\begin{equation}
\frac{d^{2} \phi}{d t^{2}}+ \frac{1}{ RC}\frac{d \phi}{d t}+
\frac{c^2}{LC}\phi +\frac{2e J_{0}}{\hbar C}\sin \phi=-\frac{2 e
c}{\hbar LC}\Phi \label{New:Transformed:Kirchhoff2}
\end{equation}
Magnetic flux  is determined by the external magnetic field $H$ and by
a cross section of a split ring with radius $a$:
\begin{equation}
\Phi=\int_{S}H ds  \approx \pi a^{2}H. \label{flux}
\end{equation}
The modulus of magnetization vector can be defined  as
\begin{equation}
M = \rho \frac{\pi a^2}{c}J, \label{Magnetization}
\end{equation}
where $\rho$ is the density of currents (or contours). Thus magnetic
response is  governed by the  following system
\begin{eqnarray}
\frac{d^{2} \phi}{d t^{2}}+ \frac{1}{ RC}\frac{d \phi}{d t}+
\frac{c^2}{LC}\phi +\frac{2eJ_{0}}{\hbar C}\sin \phi=-\frac{2 e c
\pi  a^{2}}{\hbar LC}H \label{SRR:JJ:8}\\
M= -\frac{\pi a^{2} \rho}{L} \left(\Phi + \frac{\hbar c}{2e}\phi \right)
\label{Magnetization}
\end{eqnarray}

Flux and magnetic field in~(\ref{SRR:JJ:8})
and~(\ref{Magnetization}) can be expressed via the external magnetic
field using equation~(\ref{SRR:JJ:6amod}).  Thus equations
~(\ref{SRR:JJ:8}) and~(\ref{Magnetization}) can be rewritten in the
following form
\begin{eqnarray}
\frac{\partial ^2 \phi}{\partial t^2} &+& \Gamma\frac{\partial
\phi}{\partial t} + \omega_T^2 \phi + \vartheta \sin \phi= \notag \\
&= &-\frac{2ec (\pi a^2)}{CL\hbar }\left[H^{(in)} - \frac{4 \pi
\sqrt{\varepsilon}l}{c} \frac{\partial M^{(s)}}{\partial t} \right], \label{SRR:JJ:8a} \\
M^{(s)} &=& -\frac{\pi a^2 \rho}{L} \left( \frac{\hbar c}{2e}\phi
+  \pi a^2 H^{(in)} \right).  \label{SRR:JJ:8b}
\end{eqnarray}
Here $\omega_T$ is Thomson frequency ($\omega_T^2= c^2/CL$),
coefficient $\Gamma = 1/CR$ describes dissipation, and $\vartheta =
(2eJ_0)/(C\hbar)$ determines strength of  nonlinearity. The coefficient
 $\hbar / 2e$ can be represented via  the quantum of magnetic flux
$\Phi_0 =\pi \hbar c/e $.
Let us introduce dimensionless variables using following scaling:
\begin{eqnarray*}
&&H^{in}(t)=\frac{\hbar c}{2 e \pi a^2}h(t/\tau_0),~~\tau_0 = \omega_{T}^{-1},\\
&&M(t)=\frac{\hbar c \pi a^{2} \rho}{2 e L}m(t/\tau_0).
\end{eqnarray*}
The system of equations (\ref{SRR:JJ:8a}) and (\ref{SRR:JJ:8b}) in new variables reads
\begin{eqnarray}
&&\frac{\partial ^2 \phi}{\partial \tau^2} + \gamma\frac{\partial
\phi}{\partial \tau} +  \phi + \kappa \sin \phi= - \left(h - \delta \psi \right),\notag \\
&&\psi=-\left( \frac{\partial \phi}{\partial \tau} +\frac{\partial
h}{\partial \tau} \right), \label{SRR:JJ:8ñ}
\end{eqnarray}
here $\tau = \omega_{T} t$, $\gamma = \Gamma/\omega_T$, $\kappa =
\vartheta/\omega_{T}^{2}$ and $$\delta = l\left[\frac{4 \pi
\sqrt{\epsilon} \rho (\pi a^2)^2 \omega_T}{cL}\right]=l\left[\frac{4
\pi \sqrt{\epsilon} \rho (\pi a^2)^2 }{L\sqrt{LC}}\right].$$

This model describes the magnetic response of a thin film with a diluted
concentration of split rings containing Josephson junctions.
Interaction of split rings in this case occurs only via the external
electromagnetic field.   Effects of near neighbor interaction
between  split rings were considered in the literature and can be
found
in~\cite{Lazarides:06,Lazarides:07,Lazarides:08a,Lazarides:08b,Lazarides:09}.
Work presented in~\cite{Lazarides:06,Lazarides:08a,Lazarides:08b}
considers   interacting split rings without Josephson junctions.
Dense arrangement of split rings with Josephson junctions, which
requires consideration of near neighbor interactions is
presented in~\cite{Lazarides:07,Lazarides:09}. In these papers the
external force acting on an array of split rings was determined only
by external magnetic field. Based on this work we derive a
generalization of the model presented
in~\cite{Lazarides:07,Lazarides:09}. Our equations   take into
account the influence of an additional magnetic field due to induced
magnetization of a split rings. This additional effect is accounted for
by the second term in the equation~(\ref{SRR:JJ:6amod}). This effect
results in an increase of the energy dissipation rate due to field
radiation  from the film. An additional dissipation changes  the
magnetization relaxation process  to the equilibrium states.
Additionally, the  effect of induced magnetization results in a strong
dependance of relaxation process on the frequency of external field.

\subsection{Impact of dynamical magnetic self inductance}

\begin{figure}[h!]
  \centering \includegraphics[width=2in]{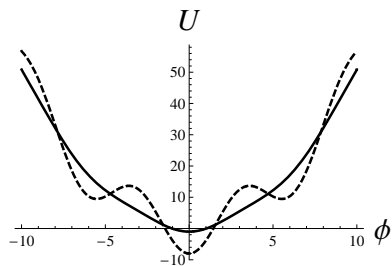}
    \caption{Graph illustrates the dependance of potential $U(\phi)=\phi^{2}/2
-\kappa \cos{\phi}$ on $\phi$. The potential can have several minima.
Solid line illustrates one minimum potential ($\kappa = 1$), dashed
line illustrates three minimum potential ($\kappa =
8$).}\label{potential}
\end{figure}

\begin{figure}[h!]
  \centering \includegraphics[width=1.5in]{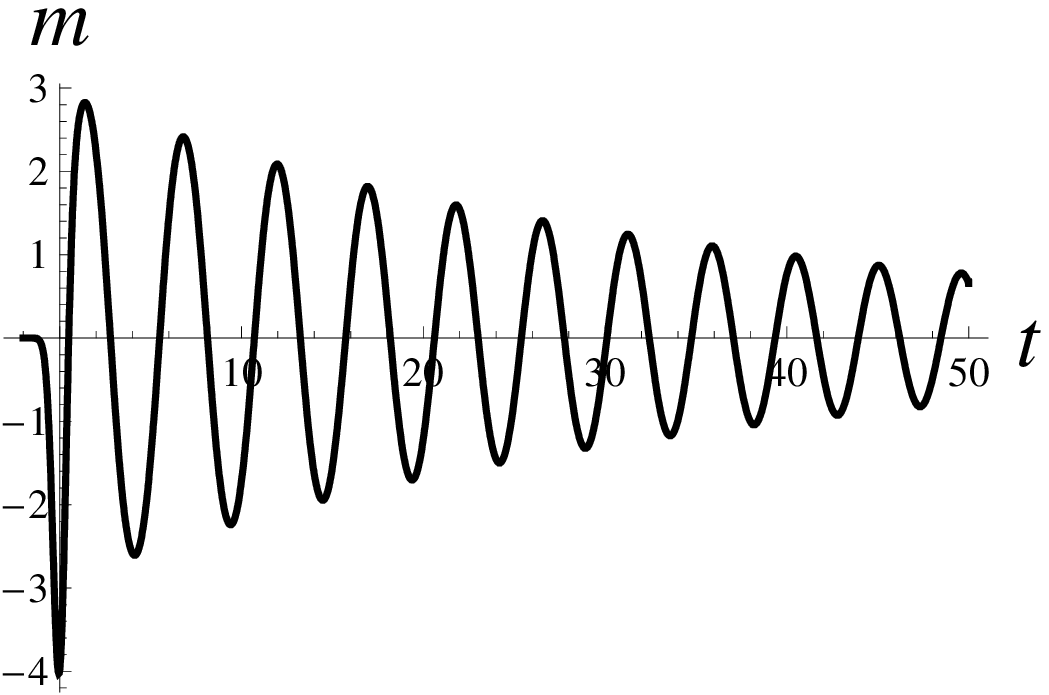}
             \includegraphics[width=1.5in]{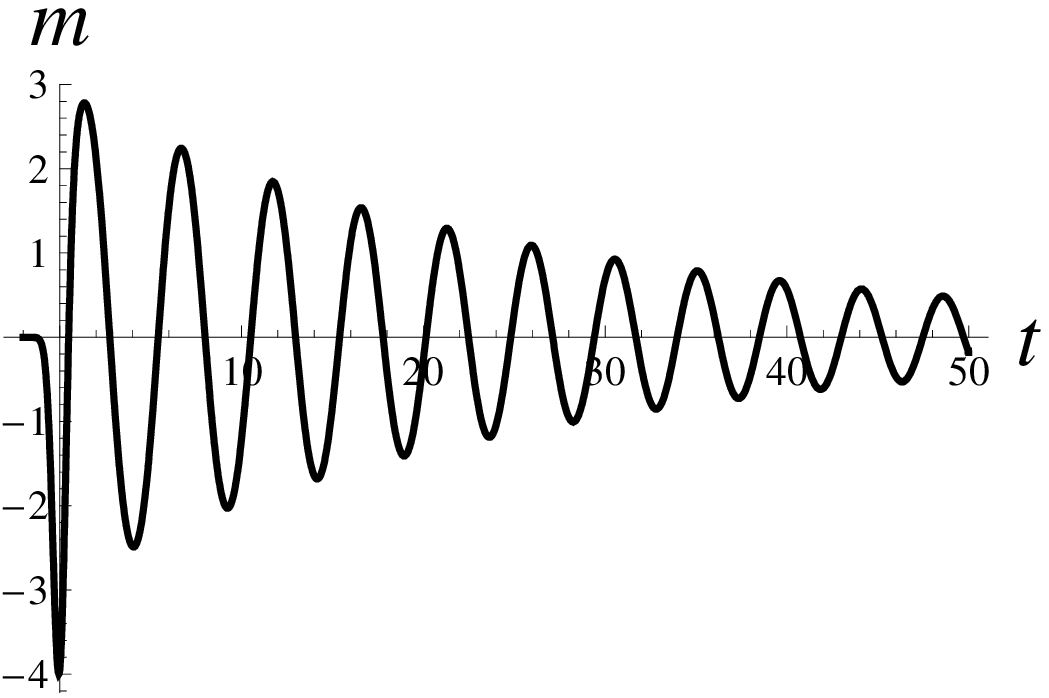}
    \caption{ Magnetization relaxation in the case when self induced magnetization
of a split ring is not taken into account - left figure and is taken
into account -right figure.  Magnetization is excited by  by
incident spike  of the gaussian shape $h(t)=a
\exp\left[-(t/t_0)\right]$, $a=4.5$ and $t_0 =0.5$. Evolution takes
place in a single  minimum potential   $U(\phi)=\phi^{2}/2 -\
\cos{\phi}$  ($\kappa = 1$ - solid line on Fig.~\ref{potential}).
}\label{magnetization:relaxation:magnetization_one_min}
\end{figure}

\begin{figure}[h!]
  \centering \includegraphics[width=1.6in]{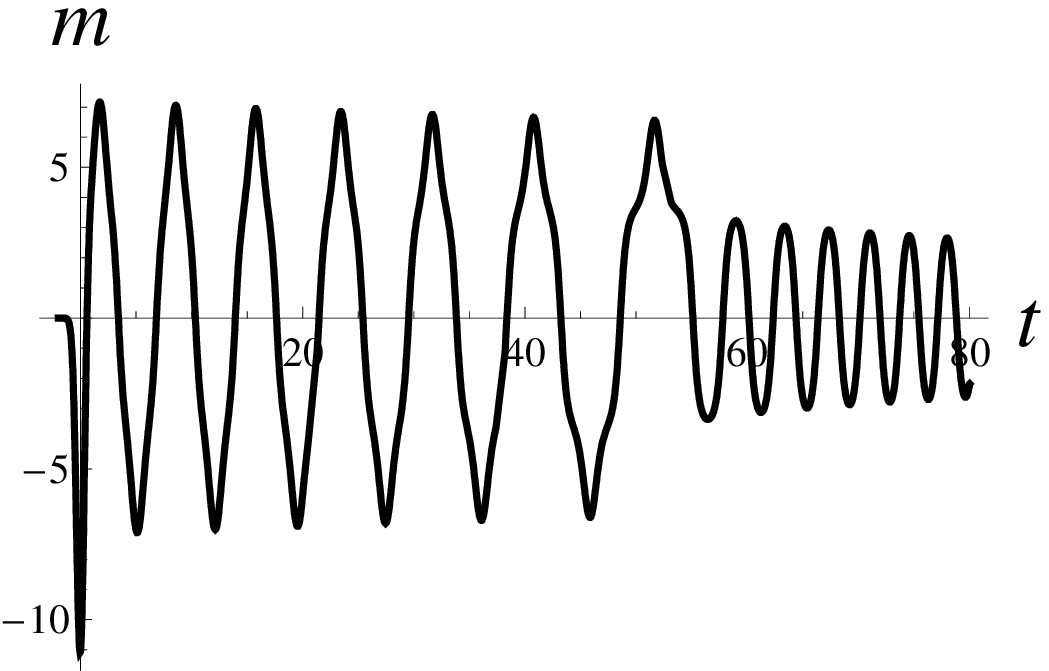}
              \includegraphics[width=1.6in]{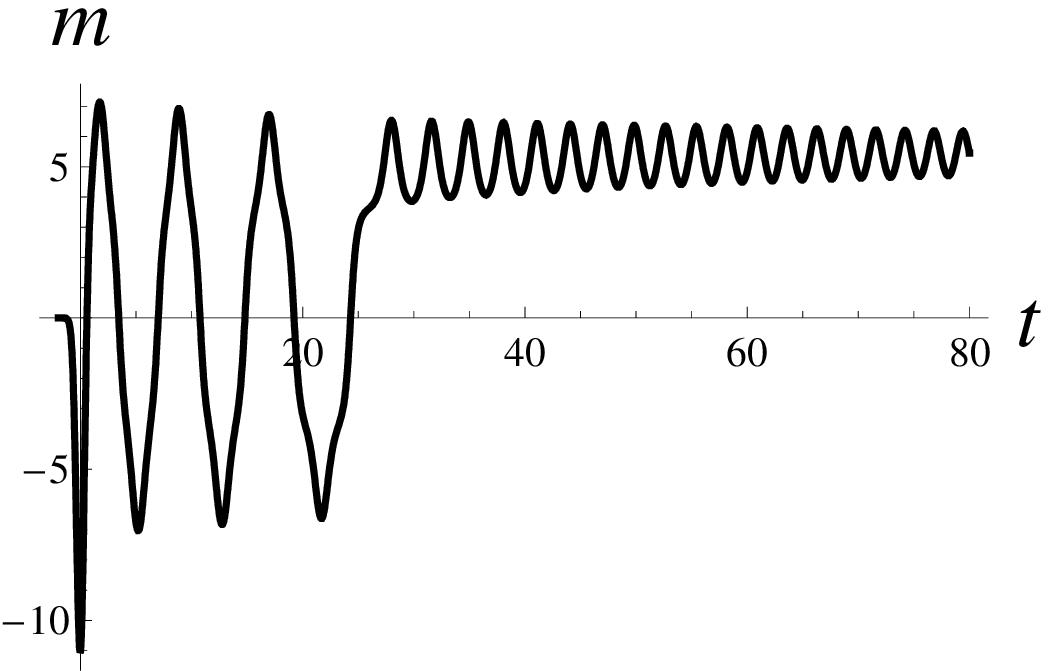}
    \caption{Magnetization relaxation in the case when self induced magnetization of a
            split ring is not taken into account - left figure and is taken into account -right figure.
            Magnetization is excited by  by incident spike  of the gaussian shape $h(t)=a \exp\left[-(t/t_0)\right]$,
            $a=12.1$ and $t_0 =0.5$. Evolution takes place in a two
            minima potential   $U(\phi)=\phi^{2}/2 -\kappa \cos{\phi}$
             ($\kappa = 8$ - dashed line on Fig.~\ref{potential}).  }
             \label{magnetization:relaxation:three:minima}
\end{figure}

Evolution of a magnetic field described by the
equations~(\ref{SRR:JJ:8ñ}) can be illustrated in terms of the dynamics
of a Newtonian particle in the potential
\[
U(\phi) =\frac{1}{2}\phi^{2}-\kappa \cos\phi.
\]
This potential can have different numbers of  minima (stable
stationary points). The number of such  minima is determined by the value of
$\kappa$.  Fig.~\ref{potential} illustrates a potential with one
minimum (solid line), where $\kappa =1$, and with three minima (dashed
line), where $\kappa = 8$. Two maxima corresponding to the last case
describe unstable states. First we consider potential with one
minimum.
From Equations~(\ref{SRR:JJ:8ñ}) it follows that taking into account the
effect of self induced magnetization results in the increase of
damping factor $\gamma \rightarrow \gamma + \delta$. This effect is
illustrated in
Fig.~\ref{magnetization:relaxation:magnetization_one_min}. The left
subfigure in
Fig.~\ref{magnetization:relaxation:magnetization_one_min} shows the
relaxation of magnetization without taking into account self induced
magnetization, and the right subfigure shows relaxation in the presence of self
induced magnetization. The relaxation process, which is shown in the
right subfigure is faster than relaxation in the left figure.
Magnetization in  this case is excited by an external magnetic field
of a gaussian shape $h(t)=a \exp\left[-(t/t_0)\right]$, $a=4.5$ and
$t_0 =0.5$. Parameters of the equations~(\ref{SRR:JJ:8ñ}) for this
example  have been chosen as follows: $\kappa = 1$, $\gamma =0.05$,
$\delta = 0.02$
Second, we consider the evolution of magnetization in the case of a three
minimum potential. The difference in the evolution of magnetization between the
two cases with and without self induced magnetization is more
dramatic in a three minimum potential. There are four oscillatory
regimes corresponding to oscillations around the central minimum, around
the two side minima and  when oscillations are taking place above two
local maxima of the potential (see Fig.~\ref{potential}). These
types of oscillations have different frequencies and, in the last
case, the period of oscillations is largest and the amplitude  is limited
from below. Fig.~\ref{magnetization:relaxation:three:minima} shows the
dynamics of magnetization with and without self induced
magnetization for the same  values of parameters and external
magnetic field.
The initially excited oscillations corresponding to the trajectories of
Newtonian particles  above local minima of the potential are
decaying and eventually switching to faster and lower amplitude
oscillations around one of the minima.  The moment of switching and
asymptotic  stable state is highly sensitive to the presence of self
induced magnetization. In this particular case, which is  illustrated
in Fig.~\ref{magnetization:relaxation:three:minima}, switching to
oscillations around the central minimum occurs  after seven cycles of
large amplitude oscillations - left subfigure.  The presence of an
additional decay in relaxation of magnetization due to self induced
magnetization, which results in excessive radiation, results in the
switching to oscillations around right minima only after three and a
half oscillatory cycles - right subfigure. In both cases
oscillations are excited by an external spike of the magnetic field
$h(t)=a \exp\left[-(t/t_0)\right]$, $a=12.1$ and $t_0 =0.5$, where
$\kappa = 8$.

\subsection{Multistability of magnetization}

\begin{figure}[h!]
  \centering \includegraphics[width=2in]{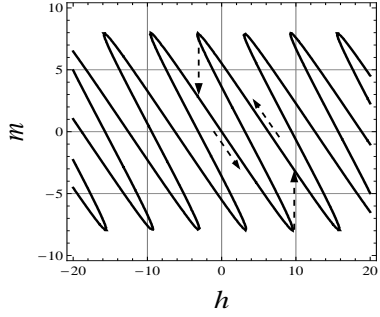}
    \caption{Graph illustrates multistability of magnetization $m$ versus
    value of stationary magnetic field $h$. Parameters of three minima potential
    are $\kappa = 8$ - dashed line on Fig.~\ref{potential}). }
    \label{multistability}
\end{figure}

The mirrorless optical bistability has been predicted in a bulk
\cite{BBE:86} and in the thin film  \cite{Bash:88} of dipole-dipole
interacting two-level atoms. We can expect this effect in the case
under consideration too.  In the case of stationary fields,
system~(\ref{SRR:JJ:8ñ}) reads as
\begin{eqnarray}
&& (m+h) + \kappa \sin (m+h)= h,~~~h=const, \label{stationary:case}
\end{eqnarray}
This is an implicit equation for $\phi$ as function of
$h$. The general solution of~(\ref{stationary:case}) can have several
roots. For a three minimum potential, the solution is illustrated in
Fig.~\ref{multistability}.
The presence of three stable and two unstable stationary solutions leads
to the multivalued nature of the function $\phi(h)$, which results in
multistability of magnetization. This function consists of stable
and unstable branches, which lead to hierarchical set of hysteresis
loops. The simplest first order hysteresis loop is shown in
Fig.~\ref{multistability}.

\section{Near neighbor interaction interaction  between split rings
}\label{SSR:JJ:2}

Following the work in~\cite{Lazarides:06} let us take into account the
magnetic dipole-dipole interaction between neighboring SRRs
characterized by cross-inductance  $L_{cross}$. Note that
cross-inductance   decays as the cube of the distance. Let us
restrict ourself to  only nearest neighbor SRR interactions.

The voltage on the $n$th junction is expressed via phase $\phi_n$ as
\begin{equation}
\label{SRR:JJ:9m} U_{n,c} = \frac{\hbar}{2 e}\frac{\partial
\phi_n}{\partial t}
\end{equation}
The voltage on the junction is equal to $U_{n,C}$ and is also equal to the
voltage $U_{n,R}$. Thus the normal (i.e., non
superfluid) current thought the junction is $U_{n,c}/R$.

According to Kirchhoff law,  voltage on the ring is equal to the sum of
acting  electromotive forces. Therefore, the  first Kirchhoff's
equation has following form:
\begin{equation}
\label{SRR:JJ:11} U_{n,c} = -\frac{L}{c^2}\frac{\partial
I_n}{\partial t}-\frac{L_{cross}}{c^2}\left(\frac{\partial
I_{n-1}}{\partial t} + \frac{\partial I_{n+1}}{\partial t}\right)
-\mathcal{E}_n
\end{equation}
Here $\mathcal{E}_n=-c^{-1}d\Phi_n/dt$ is electromotive force induction generated
by high frequency magnetic flux acting on $n$th ring, $I_{n \pm 1}$ is the
current in the $(n\pm 1)$th ring.

The second Kirchhoff equation corresponds to charge conservation:
\begin{equation}
\label{SRR:JJ:12} I_n=J_n+C_n \frac{\partial U_{n,c}}{\partial
t}+\frac{U_{n,c}}{R}
\end{equation}

Combining equation~(\ref{SRR:JJ:9m}) with~(\ref{SRR:JJ:11}) we obtain
\begin{equation}
I_n+\eta \left(I_{n-1}+I_{n+1}\right) = -\frac{c}{L} \left( \Phi_n +
\frac{c\hbar}{2e}\phi_n\right), \label{SRR:JJ:13}
\end{equation}
where $\eta = L_{cross}/L$. This equation must be complemented with
equations defining currents at each split-ring:
\begin{equation}
I_n = J_{0n}\sin\phi_n + \frac{C\hbar}{2e}\frac{\partial ^2
\phi_n}{\partial t^2}+ \frac{\hbar}{2er}\frac{\partial
\phi_n}{\partial t}. \label{SRR:JJ:14}
\end{equation}

Expression~(\ref{SRR:JJ:13}) represents a linear recurrency
relation. In the limit $N \rightarrow \infty $, we can employ the
generating function method.

\subsection{Calculation of individual currents in a split rings}

Let us introduce notation
\[
F_n=-\frac{c}{L} \left( \Phi_n + \frac{c\hbar}{2e}\phi_n\right).
\]
and rewrite equation (\ref{SRR:JJ:13}) as
\begin{equation}
I_n+\eta \left(I_{n-1}+I_{n+1}\right)=F_n, \label{SRR:JJ:15}
\end{equation}
Introducing the generating function
\[
P(y)=\sum \limits_{n=-\infty}^{+\infty} I_n e^{iny},
\]
defining the function $F(y)$
\begin{equation}
F(y)= \sum \limits_{n=-\infty}^{+\infty} F_n e^{iny}\label{Definition:F}
\end{equation}
and using  equation (\ref{SRR:JJ:15}) we can find relation between $P(y)$ and $F(y)$
\begin{equation}
P(y)=\frac{F(y)}{1+2\eta \cos y}.  \label{SRR:JJ:16}
\end{equation}
This expression can be represented as a power series
\begin{equation*}
P(y)=\sum_{m=0}^{\infty} (-2\eta)^m \cos^m (y) F(y).
\end{equation*}
Function  $\cos^m (y)$ can be expressed  as
\begin{eqnarray*}
\cos^m (y) &=& 2^{-m} e^{imy}\left( 1+e^{-2iy}\right)^{m} = \notag \\
&=& 2^{-m} e^{imy}\sum_{p=0}^{m}\binom {m}{p}e^{-2ipy},
\end{eqnarray*}
then the generating function has the form:
\begin{equation*}
P(y)=\sum_{m=0}^{\infty} (-2\eta)^m \sum_{p=0}^{m}\binom
{m}{p}e^{-iy(m-2p)} F(y).
\end{equation*}
The current in the $n$-th ring is determined by the formula
\[
I_n = (2\pi)^{-1}\int_{-\infty}^{+\infty} P(y) e^{iny}d y.
\]
Using the expression for $F(y)$~(\ref {Definition:F}) we can find that
\begin{equation}
I_n(t) = \sum_{m=0}^{\infty}(-\eta)^m\sum_{p=0}^{m}\binom {m}{p}
F_{n+m-2p}(t). \label{SRR:JJ:17}
\end{equation}
The expression for $I_n$ can be simplified taking into account smallness
of the parameter $\eta \ll 1$. Up to the second order of $\eta $, the
expression for $I_n(t)$ has the form
\begin{eqnarray}
I_n(t)&\approx &F_n(t) - \eta \left(F_{n+1}(t)+F_{n-1}(t) \right) + \notag \\
&+&\eta^2 \left(F_{n+2}(t)+2F_{n}(t)+F_{n-2}(t) \right).
\label{SRR:JJ:18}
\end{eqnarray}

\subsection{Continuum chain approximation}

In the continuum approximation we can introduce a new variable
$y=nl_r$, where  $l_r$ is the distance between neighboring
rings, and represent  $ F_{n \pm 1}(t)$ as
$$ F_{n \pm 1}(t) \approx  F(y)(t) \pm l_r
\frac{\partial F}{ \partial y}+ \frac{ l_r^2}{2 } \frac{\partial^2
F}{ \partial y^2}. $$
Limiting ourself to the first order  expansion with respect to $\eta$ we obtain from  (\ref{SRR:JJ:18})
\begin{eqnarray}
I(y,t) &\approx &(1-2\eta)F(y,t) - \eta l_r^2 \frac{\partial^2 }{
\partial y^2}F(y,t) \notag \\
&\approx& F(y,t) - \eta l_r^2 \frac{\partial^2 }{
\partial y^2}F(y,t). \label{SRR:JJ:18m}
\end{eqnarray}

In our model $F(y,t)$ was defined as
\[
F(y,t)=-\frac{c}{L} \left( \Phi (y,t) +
\frac{c\hbar}{2e}\phi(y,t)\right)
\]
Since  the magnetic flux through ring is greater than the  flux through the
junction (gap),  we can use the  approximate expression
$\Phi(y,t) \approx (\pi a^2)H(y,t)$. Substitution of this expression   in
(\ref{SRR:JJ:14}) leads to
\begin{eqnarray}
J_{0n}\sin\phi_n &+& \frac{C\hbar}{2e}\frac{\partial ^2
\phi_n}{\partial t^2}+ \frac{\hbar}{2eR}\frac{\partial
\phi_n}{\partial t} = \notag \\
&=& -\frac{c}{L} \left( \Phi (y,t) + \frac{c\hbar}{2e}\phi(y,t)\right)+ \notag \\
&+&\frac{\eta l_r^2 c}{L} \left( \frac{\partial^2 \Phi}{\partial
y^2} + \frac{c\hbar}{2e}\frac{\partial ^2\phi}{\partial y^2}\right).
 \label{SRR:JJ:19}
\end{eqnarray}
Using dimensionless  variables defined above, the resulting system of
equations can be written as
\begin{eqnarray}
\frac{\partial ^2 \phi}{\partial \tau^2} &-&
\frac{\partial ^2 \phi}{\partial \xi^2} + \gamma\frac{\partial
\phi}{\partial \tau} +  \phi + \kappa \sin \phi = \notag
\\
&=&- \left(h - \delta\psi \right)+ \frac{\partial ^2}{\partial
\xi^2}\left(h - \delta\psi \right),
\label{SRR:JJ:20a} \\
\psi &=&-\left( \frac{\partial
\phi}{\partial \tau} +\frac{\partial
h}{\partial \tau} \right). \label{SRR:JJ:20b}
\end{eqnarray}
Here we  introduce dimensionless spatial variable $\xi
=y/(l_r \sqrt{\eta})$.

In case of plane electromagnetic wave propagating along $x$, dependence of $h$ on $y$ vanishes. Spatial derivatives in the equation (\ref{SRR:JJ:20b}) disappear and equation describes homogeneous dynamics of the chain as function of time. Temporal-spatial chain dynamics takes place when the chain is excited by an electromagnetic beam with spatial distribution of intensity across the beam. In the general case, excitation of surface waves requires matching of the surface wave vector with tangential component of the vector of the incident wave. Equation~(\ref{SRR:JJ:20a}) demonstrates that sharp spatial gradients across the beam act as an external force for oscillations along the chain.

\section{Small-amplitude approximation in SRRs}

In the linear approximation equation (\ref{SRR:JJ:8a}) reads
\begin{equation}
\label{SRR:JJ:lin} \frac{\partial ^2 \phi}{\partial t^2} +
\frac{1}{Cr}\frac{\partial \phi}{\partial t} + \Omega ^2 \phi
=-\frac{2ec}{CL\hbar }\Phi (t)
\end{equation}
where $\Omega$ is renormalized Thompson frequency  $\Omega^2 =
(LC/c^2)^{-1}+ (2eJ_0/C\hbar)$. The Fourier components of current
are represented as
\begin{equation*}
I(\omega)=\frac{\pi a^2 c}{L}\left(\frac{\omega_T^2}{\Omega^2 -
\omega^2 - i\Gamma \omega}-1 \right) H(\omega).
\end{equation*}

Fourier components of magnetic induction $B=H+4\pi M$ are expressed
through magnetic permittivity $B=\mu (\omega)H$, which leads to
\begin{equation*}
\mu(\omega)=1+4\pi n_m \frac{(\pi
a^2)^2}{L}\left(\frac{\omega_T^2}{\Omega^2 - \omega^2 - i\Gamma
\omega}-1 \right)
\end{equation*}

Since equation (\ref{SRR:JJ:8a}) is nonlinear, the magnetic response is described by a nonlinear coupling between field $H$ and magnetization $M$. The total magnetization can be represented as sum of a linear part $M_{lin}$ and nonlinear part $M_{nl}$. Taking into account the definition of
magnetization~(\ref{SRR:JJ:8b})  we obtain  following expressions
\[
M_{lin} = -n_m\frac{\pi a^2}{L}\left( \pi a^2 H +\frac{\hbar
c}{2e}\phi^{(0)}\right),
\]
\[
M_{nl} = -n_m\frac{\pi a^2}{L}\frac{\hbar c}{2e}\phi^{(1)},
\]
where $\phi^{(0)}$ is solution of the linear equation, and $\phi^{(1)}$
is correction to solution of the linear equation (\ref{SRR:JJ:lin}).

Expansion of  sine-function up to   cubic  term $\phi^3$),
transforms  (\ref{SRR:JJ:8a}) into the Duffing equation:
\begin{equation}
\label{eqJJ:14} \frac{\partial ^2 \phi}{\partial t^2} + \Gamma
\frac{\partial \phi}{\partial t} + \Omega ^2 \phi + \vartheta \phi^3
=-\frac{2ec}{CL\hbar }\Phi (t)
\end{equation}
where $ \vartheta = 2eJ_0/6c\hbar $ is parameter characterizing
nonlinear response of Duffing oscillator. Let us consider only the first
correction term for solution of (\ref{SRR:JJ:8a}). Substituting  $\phi
= \phi^{(0)}+\phi^{(1)}$ into (\ref{SRR:JJ:8a}) and collecting  terms
of same order we obtain:
\begin{eqnarray}
&& \frac{\partial ^2 \phi^{(0)}}{\partial t^2} + \Gamma
\frac{\partial \phi^{(0)}}{\partial t} + \Omega ^2 \phi^{(0)}
=-\frac{2ec}{CL\hbar }\Phi (t)
 \label{eqJJ:15a}\\
&& \frac{\partial ^2 \phi^{(1)}}{\partial t^2} + \Gamma
\frac{\partial \phi^{(1)}}{\partial t} + \Omega ^2 \phi^{(1)} =
-\vartheta \phi^{(0)3}.
 \label{eqJJ:15b}
\end{eqnarray}

Solution  of (\ref{eqJJ:15a}) gives a zero order approximation for $\phi$:
\begin{equation}
\label{eqJJ:16} \phi^{(0)}(\omega)= - \delta_m
\mathfrak{L}(\omega)H(\omega),
\end{equation}
where $$\delta_m=\frac{2\pi c e a^2}{CL\hbar} = \frac{2\pi
\omega^2_T}{\Phi_0}(\pi a^2)$$ is a coupling parameter. Function
\[
\mathfrak{L}(\omega)=\frac{1}{\Omega^2 - \omega^2- i \omega \Gamma}
\]
is  the standard \textit{Lorentzian} function. For harmonic wave with
carrier frequency  $\omega_0$, we have
\[
H(\omega) = H_0 \delta (\omega-\omega_0) +H_0^{*} \delta
(\omega+\omega_0).
\]
In this case
\[ \phi^{(0)}(\omega)= - \delta_m \left[
\mathfrak{L}(\omega)H_0 \delta (\omega-\omega_0)
+\mathfrak{L}(-\omega)H_0^{*} \delta (\omega+\omega_0)\right],
\]
and
\begin{equation}
\label{eqJJ:17} \phi^{(0)}(t)= - \delta_m \left[
\mathfrak{L}(\omega_0)H_0 e^{-i\omega_0 t}
+\mathfrak{L}(-\omega_0)H_0^{*} e^{i\omega_0 t} \right].
\end{equation}

The next step is to substitute expression for   $\phi^{(0)}(t)$ into the  right hand
part of the equation  (\ref{eqJJ:15b}):
\begin{eqnarray*}
\vartheta \phi^{(0)3}(t) &=&
\tilde{\alpha}_1(\omega_0)H_0^3 e^{-3i\omega_0} +\\
&+&\tilde{\alpha}_2(\omega_0)|H_0|^2 H_0 e^{-i\omega_0} + c.c..
\end{eqnarray*}
Here $\tilde{\alpha}_1(\omega_0)$ and $\tilde{\alpha}_2(\omega_0)$ are defined as follows:
\begin{eqnarray*}
&&\tilde{\alpha}_1(\omega_0) = \vartheta
\delta_m^3\mathfrak{L}(\omega_0)\mathfrak{L}(\omega_0)\mathfrak{L}(\omega_0),\\
&&\tilde{\alpha}_2(\omega_0) = 3\vartheta
\delta_m^3\mathfrak{L}(\omega_0)\mathfrak{L}(\omega_0)\mathfrak{L}(-\omega_0).
\end{eqnarray*}

From  (\ref{eqJJ:15b}), taking account of the results obtained
above, we can find  $\phi^{(1)}(\omega)$
\begin{eqnarray*}
\phi^{(1)}(\omega) &=&
-\tilde{\alpha}_1(\omega_0)\mathfrak{L}(\omega_0)H_0^3 \delta(\omega-3\omega_0) -\\
&&-\tilde{\alpha}_2(\omega_0)\mathfrak{L}(\omega_0)|H_0|^2 H_0
\delta(\omega- \omega_0) + c.c..
\end{eqnarray*}

Having $\phi^{(0)}(\omega)$ and $\phi^{(1)}(\omega)$ we can write
expressions for the linear and nonlinear parts of magnetization:
\begin{eqnarray*}
&&M_{lin}(\omega) = n_m\frac{\pi a^2
}{L}\left[\omega_T^2\mathfrak{L}(\omega) -1 \right]H(\omega),\\
&&M_{nl}(\omega) = n_m\frac{\pi a^2\hbar c}{2eL}\left[\alpha^{(3)}_1H_0^3
\delta(\omega-3\omega_0) \right. + \\
&&~~~~~~+\left. \alpha^{(3)}_2|H_0|^2 H_0 \delta(\omega- \omega_0) +
c.c.\right].
\end{eqnarray*}
Here we use notations for nonlinear magnetic susceptibilities of
third order
\begin{eqnarray*}
&&\alpha^{(3)}_1 = \alpha^{(3)}(3\omega
;\omega_0,\omega_0,\omega_0)=\vartheta
\delta_m^3\left[\mathfrak{L}(\omega_0)\right]^3 \mathfrak{L}(3\omega_0),\\
&&\alpha^{(3)}_2 =
\alpha^{(3)}(\omega_0;\omega,\omega_0,-\omega_0)=3\vartheta
\delta_m^3\left[\mathfrak{L}(\omega_0)\right]^3\mathfrak{L}(-\omega_0).
\end{eqnarray*}
The first term in the expression for $M_{nl}(\omega)$ describes
process of  third harmonic generation, second term in this
expression describes phenomena of Kerr's self-modulation.

\section{Conclusion}

We derived equations describing electromagnetic pulse interaction with thin films containing split rings with Joshephson junctions, taking into account  dynamical self-inductance in the split rings, and analyzed the impact of dynamical self-inductance on the transmitted and reflected wave.  We demonstrated an increase of magnetization relaxation rate   due to dynamical self-inductance.  This additional damping can results in significant  change in the evolution dynamics of film magnetization. We also derived an equation describing the interaction of a chain of split-rings containing  JJ with an electromagnetic field and  showed that    the  current in a particular ring is defined by the magnetic fluxes through all other rings in a chain.  The continuous limit  transforms  this model into sin-Gordon  type of equation. This equation is driven by an external force that  contains second spatial derivatives of the incident field. Presence of the field spatial derivatives   suggests a mechanism for the excitation of  magnetization waves along the chain by normally incident field. We found analytic expression for  nonlinear magnetic susceptibility of such  films  in the limit of weak amplitudes, describing  third harmonic generation and self modulation due to high frequency Kerr effect.

\section*{Acknowledgment}

We would like to thank  Alexey V. Ustinov  for enlightening discussions and Matthew F. Pennybacker for valuable help during preparation of this paper. A.I.M appreciates  support and hospitality of the University of Arizona Department of Mathematics  during his work on this manuscript. This work was partially supported by NSF (grant DMS-0509589), ARO-MURI award 50342-PH-MUR  and State of Arizona (Proposition 301), RFBR (grant No. 09-02-00701-a).


\begin{thebibliography}{99}



\bibitem{Pendry:04} \textit{J.B. Pendry}, Negative refraction,
 Contemporary Physics. \textbf{45}, 191-202 (2004)

\bibitem{Veselago:06} \textit{V. Veselago, L.
Braginsky, V. Shklover, Ch. Hafner}, Negative Refractive Index
Materials  J. Computational and Theoretical Nanoscience. \textbf{3},
1-30 (2006)


\bibitem{Agranovich:06} \textit{V M Agranovich, Yu N Gartstein}, Spatial dispersion and negative refraction of light,
Physics--Uspekhi \textbf{49} (10) 1029-1044 (2006)

\bibitem{Maim:Gabi:07} \textit{A.I. Maimistov, I.R. Gabitov}, Nonlinear optical effects
in artificial materials, Eur. Phys. J. Special Topics. \textbf{147},(1), 265-286 (2007)

\bibitem{Maim:Gabi:Lit:08} \textit{A.I. Maimistov, I.R. Gabitov, N.M. Litchinitser}, Solitary
Waves in a Nonlinear Oppositely Directed Coupler, Optics and
Spectroscopy, \textbf{104}, 253–257 (2008),

\bibitem{Kaz:Maim:Ozh:09} \textit{ E.V. Kazantseva, A.I.
Maimistov, S.S. Ozhenko}, Solitary electromagnetic waves propagation
in the asymmetric oppositely-directed coupler, arXiv:0904.4035v1
[nlin.PS]

\bibitem{Rautian:97} \textit{S. G. Rautian}, Nonlinear saturation spectroscopy of the degenerate
electron gas in spherical metallic particles, JETP \textbf{85},
451–461 (1997).

\bibitem{Manfredi:97} \textit{N. Crouseilles, P.-A. Hervieux, G. Manfredi},
Quantum hydrodynamic model for the nonlinear electron dynamics in
thin metal films, Phys.Rev. B \textbf{78}, 155412 (2008)


\bibitem{Lapine:03} \textit{M. Lapine, M. Gorkunov, K. H. Ringhofer}, Nonlinearity of a
metamaterial arising from diode insertions into resonant conductive
elements, Phys.Rev. E \textbf{67}, 065601–4 (2003).

\bibitem{Shadrivov:06} \textit{I.V. Shadrivov, S.K. Morrison, Yu.S. Kivshar}, Tunable
split-ring resonators for nonlinear negative-index metamaterials,
Optics Express \textbf{14}, 9344-9349 (2006)


\bibitem{Hansen:84} \textit{J. Bindslev Hansen, P.E. Lindelof},
Static and dynamic interactions between Josephson junctions,
Rev.Mod.Phys. \textbf{56}, 431-459 (1984)

\bibitem{Lazarides:06} \textit{N. Lazarides, M. Eleftheriou, and G. P.
Tsironis1}, Discrete Breathers in Nonlinear Magnetic Metamaterials,
Phys.Rev.Lett. \textbf{97}, 157406 (2006)

\bibitem{Lazarides:07} \textit{N. Lazarides, G. P.
Tsironis, M. Eleftheriou}, Dissipative discrete breathers in rf
SQUID metamaterials, arXiv:0712.0719v1 [nlin.PS]

\bibitem{Lazarides:09} \textit{G.P. Tsironis1, N. Lazarides1, and M. Eleftheriou},
Dissipative Breathers in rf SQUID Metamaterials, PIERS Proceedings,
pp. 52-65, Beijing, China, March 23–27, 2009


\bibitem{GabMaimSPIE08} \textit{I.R. Gabitov and A.I. Maimistov}, Nonlinear
response of metamaterials based on Josephson junction arrays,
Proceedings of SPIE  \textbf{7029}, pp. 47,  SPIE Optics+Photonics
2008,  Metamaterials: Fundamentals and Applications, San Diego
August 10-13, 2008

\bibitem{Rupasov:82} \textit{V.I. Rupasov, and V.I. Yudson}, On the boundary problems of nonlinear optics of resonant media,
Sov.J.Quantum Electron. \textbf{12}, 415-419 (1982).

\bibitem{Rupasov:87} \textit{V.I. Rupasov, and V.I. Yudson}, Nonlinear resonant optics of thin films:
the method of inverse scattering transformation, Sov.Phys. JETP \textbf{66}, 282-285(1987).

\bibitem{BBE:86} \textit{Y. Ben-Aryeh, C. M. Bowden, and J. C. Englund}, Phys.Rev. A
\textbf{34},  3917 - 3926 (1986)

\bibitem{Bash:88} \textit{A.M. Basharov}, Thin film of resonant atoms:
a simple model of optical bistability and self-pulsation, Sov.Phys.
JETP \textbf{67},  1741-1744 (1988).


\bibitem{BMTZ91} \textit{Benedict M.G., Malyshev V.A., Trifonov E.D., Zaitsev
A.I.}, Reflection and transmission of ultrashort light pulses
through a thin resonance medium: local fields effects, Phys.Rev.
\textbf{A43}, 3845-3853 (1991)

\bibitem{Vanagas:98} \textit{E.Vanagas, and A.I. Maimistov},
The reflection of ultrashort light pulses from nonlinear boundary of
two dielectric media, Opt. Spektrosk. \textbf{84,} 301-306 (1998).

\bibitem{Elyutin07a} \textit{S.O. Elyutin}, Propagation of a videopulse through
a thin layer of two-level dipolar atoms, J. Phys. B: At. Mol. Opt.
Phys. \textbf{40}, 2533-2550 (2007)

\bibitem{Caputo:07} \textit{J.-G. Caputo, A.I. Maimistov
and E. V. Kazantseva}, Electromagnetically induced switching of
ferroelectric thin films, Phys.Rev. B \textbf{75}, 014113 (9 pages)
(2007).

\bibitem{Barone:82} \textit{ A. Barone and G. Paterno},
{Physics and Applications of the Josephson effect}, J. Wiley,
(1982).

\bibitem{Scott:70} \textit{A. Scott}, Active and nolinear
wave propagation in electronics, Wiley interscience, New York,
London, Sydney, Toronto, 1970.


\bibitem{Lifscits:Pita:78} \textit{E.M. Lifschitc, L.P. Pitaevskii}, Statistical Physics,
Pt.2 Theory of condenced mater, Nauka, Moscow, 1978

\bibitem{Feynman:98} \textit{R. Feynman}, Statistical Mechanics, Addison-Wesley, Reading, MA 1998.

\bibitem{Lazarides:08a} \textit{M. Eleftheriou, N.
Lazarides, and G. P. Tsironis} Magnetoinductive breathers in
metamaterials,  Phys.Rev. E \textbf{77}, 036608 (13 pages) (2008)

\bibitem{Lazarides:08b} \textit{Nikos Lazarides, George P. Tsironis, and Yuri
S. Kivshar}, Surface breathers in discrete magnetic metamaterials
Phys.Rev. E \textbf{77}, 065601(R) (4 pages) (2008)


\end{thebibliography}
\end{document}